\documentclass[a4paper,10pt]{article}
\usepackage[utf8]{inputenc}
\usepackage{amsmath}
\usepackage{slashed}
\usepackage{amssymb}
\numberwithin{equation}{section}
\usepackage[margin=1in]{geometry}

\usepackage{hyperref}
\usepackage{natbib}
\usepackage[usenames,dvipsnames,svgnames]{xcolor}
\usepackage{setspace}

\setcitestyle{square,numbers,comma}

\begin{document}
\baselineskip 24pt

\begin{center}
{\Large \bf Fermion zero modes of Denef Bates black holes}

\end{center}

\vskip .6cm
\medskip


\baselineskip=18pt

\centerline{\large \rm   Palash Dubey}

\vspace*{4.0ex}

\centerline{\large \it National Institute of Science Education and Research Bhubaneshwar,}

\centerline{\large \it  P.O. Jatni, Khurdha, 752050, Odisha, INDIA}

\vspace*{1.0ex}

\centerline{\large \it Homi Bhabha National Institute, Training School Complex, }

\centerline{\large \it  Anushakti Nagar, Mumbai 400094, INDIA}

\vspace*{4.0ex}
\centerline{E-mail: palash@niser.ac.in }

\vspace*{5.0ex}

\centerline{\bf Abstract} 
\bigskip
We study the fermion zero modes of Denef Bates black holes in $\mathcal{N}=2$ Supergravity. The goldstino zero modes are constructed explicitly and their boundary conditions, charge, and normalizability are studied. We then study if there can exist fermion zero modes that are normalizable and not gauge equivalent to the goldstino modes. We find that such modes are not possible if we want to preserve the continuity of these solutions. The non-existence of these modes shows that Denef Bates black holes must contribute to the index.
    \section{Introduction and summary}
    \label{sec:intro}
    Black holes are classical objects in general relativity that are regions of space-time enclosed by 
    an event horizon that even massless particles cannot escape. 
    But, quantum-mechanically these objects are not at zero temperature. 
    Their temperature and Bekenstein-Hawking entropy is:
    \begin{equation}\label{eq:entropy}
        T=\frac{\kappa}{2\pi},\;\;\; S=\frac{A}{4},
    \end{equation}
    where $\kappa$ is the surface gravity and $A$ is the area of the event horizon.

    Being a quantum theory of gravity string theory can be used to count the microscopic degrees of freedom of a black hole and match it with the entropy of the black hole with the relation 
    \begin{equation}
        S=\ln\Omega.
    \end{equation} 
    This has been checked successfully in string theory for supersymmetric black holes \cite{SEN_1995,Strominger_1996,Horowitz_1996,Maldacena_1996,Callan_1996,
Dijkgraaf_1997,Cardoso_2005,David_2006,Jatkar_2006,Sen_2007,Emparan_2006}. 
    In this program, one identifies a supersymmetric black hole carrying some charges and then using the charges we compute the entropy of the black hole in Supergravity which can have corrections to \ref{eq:entropy}. 
    Then we identify quantum states with the same charges in string theory. 
    This allows us to match the logarithm of the number of states with the entropy of the black hole in Supergravity. 
    The computation of the number of states can be done in a region of moduli space where gravity is weak. We compute an index which receives contributions from BPS states alone. This makes the index protected by supersymmetry, that is to say that it does not change with coupling.

    If a BPS state breaks 4n supersymmetries then there are 4n goldstino fermion zero modes. 
    Quantization of these modes produces an equal number of fermionic and bosonic states; hence the Witten index $Tr(-1)^F$ receives no contribution from these states. 
    To have a non-zero contribution from these states the helicity trace index is used. This has been reviewed in \cite{Mandal_2010}.  It is given by
    \begin{equation}
        B_{2n}=\frac{1}{2n!}Tr\left[{(-1)^F (2h)^{2n}}\right]
    \end{equation}
    where $h$ is the third component of angular momentum in the rest frame. 
    The trace is computed over states with a fixed set of charges. 
    If a state has more than 4n fermion zero modes then those states will not contribute to the index. 
    This property makes only BPS states contribute to it and also can be used to check which states contribute to the index and which do not. 
    In this paper, we are asking if Denef Bates black holes \cite{denefbates} contribute to the index. More specifically we ask the question: are there any fermion zero modes apart from the broken zero modes for  Denef Bates black holes in $\mathcal{N}=2$ Supergravity; which would make the index associated with these black holes vanish.

    In \cite{Denef:2007vg} the authors have shown that a charge $\Lambda\Gamma$ supporting single cenered Denef-Bates black holes of entropy $\sim \Lambda^2 S(\Gamma)$ also admits two-centered BPS black hole realizations whose entropy grows like $\Lambda^3$ when $\Lambda\rightarrow\infty$. 
    This shows that the entropy associated to two centered black holes is extensive and macroscopic and the two centered black holes contribute to the index.
    Hence, we expect to find no extra fermion zero modes apart from the goldstino zero modes.

    Fermion zero modes were first computed in \cite{Aichelburg:1987hy,Aichelburg:1987hx,Aichelburg:1987hz,Aichelburg:1987ia} for Majumdar Papapetrou solutions. The norm on these fermion zero modes, which is used as a central tool in this paper, was proposed in \cite{kallosh}.
    Extra fermion zero modes for the Majumdar Papapetrou black holes were studied in \cite{Dubey:2019sbf}.

    In this paper, we compute the goldstino fermion zero modes for multicentered Denef Bates black holes in $\mathcal{N}=2$ Supergravity and calculate the supercharge of these modes. 
    Then we study the normalizability and boundary conditions of these modes. 
    We find that the boundary condition for these zero charge modes is $r^2\psi\rightarrow0$ asymptotically, also they are continuous all over $\mathbb{R}^3$.
    Which shows that the norm of these zero charge modes recieves no contribution in the asymptotic limit. Then we show that to have a non-zero contribution in the near horizon limit we must have a discontinuous gravitini. Thus, we do not have a contribution to the norm from the near horizon limit also.

    The nonexistence of a zero charge and non-zero  norm mode implies that there cannot be any gauge inequivalent extra fermion zero modes in the spectrum. 
    Hence, the  Denef Bates black holes must contribute to the index.

    The paper is organized as follows. 
    In section \ref{sec:sugra} we gather the needed ingredients of $\mathcal{N}=2$ supergravity. 
    In section \ref{sec:dbbh} we introduce the Denef Bates background and also compute the Killing spinors for the background. 
    In section \ref{sec:fzms} we find the goldstino fermion zero modes, show that they satisfy the equations of motion,  find the supercharge and the boundary conditions of these solutions.
    In section \ref{sec:xfzm} we show that you need to have non-trivial zero charge solutions to have gauge inequivalent solutions; then we argue that such zero charge modes satisfying the gauge condition and equations of motion must have zero norm, and hence will not be proper states. 
    In appendix \ref{app:spinc} we calculate spin connection and other quantities needed for calculating the fermion zero modes.
    In appendix \ref{app:eomf} we compute the equations of motion of the theory.

    \section{$\mathcal{N}=2$ Supergravity}
    \label{sec:sugra}
    In this paper, we are concerned with Denef Bates black holes in the context of string theory. 
    If we reduce 10 dimensional Type IIA Supergravity (the low energy limit of type IIA string theory) on a six dimensional Calabi Yau manifold $X$, which preserves only a quarter of the supersymmetry, we get 4 dimensional $\mathcal{N}=2$ Supergravity coupled to $h^{1,1}=\text{dim}\,H^{1,1}(X)$ vector multiplets and $h^{2,1}+1$ hyper multiplets. The hyper multiplets can be completely decoupled, and we can set them to arbitrary constant values. Hence, we are interested in Denef Bates black holes as solutions to $\mathcal{N}=2$ Supergravity coupled to $n_V=h^{1,1}$ vector multiplets.

    The field content of this theory contains a graviton $g_{\mu\nu}$, two gravitini which are Rarita Schwinger fields $\psi_\mu^{(i)}$, $n_V+1$ gauge fields (containing the graviphoton) $A_\mu^I$, $2\; n_V$ gaugini $\chi_{(i)}^\alpha$ and $n_V$ complex scalar fields $z^\alpha$.
    $I$ takes values $0\cdots n_V$ and $\alpha$ takes values $1\cdots n_V$. 

    The metric we use is mostly positive. The clifford algebra in this paper is defined by
    \begin{eqnarray}
        &&\gamma^a\gamma^b+\gamma^b\gamma^a=2\eta^{ab},\;\;\;\gamma_{\mu\nu\cdots}=\gamma_{[ \mu}\gamma_{\nu}\cdots_{]},\nonumber\\
        &&\gamma_*=i\gamma_0\gamma_1\gamma_2\gamma_3.
    \end{eqnarray}
    The Chiral projection operators are
    \begin{equation}
    P_L=\frac{1}{2}\left(1+\gamma_*\right),\;\;\; P_R=\frac{1}{2}\left(1-\gamma_*\right).
    \end{equation}
    The $(i)$ indices are the rigid R-symmetry $SU(2)$ indices, they are raised and lowered by charge(or complex) conjugation. 
    \begin{equation}
        \psi^{(i)}=P_L\psi^{(i)},\;\;\;\psi_{(i)}=P_R\psi_{(i)},\;\; \psi^{(i)}=\psi_{(i)}^C.
    \end{equation}
    We also use the Levi-Civita $\varepsilon^{(ij)}$ for which the important property is that
    \begin{equation}
        \varepsilon_{(ij)}\varepsilon^{(jk)}=-\delta_i^k.
    \end{equation}
    These are the conventions and definitions of \cite{freedman}.
    \subsection{Special K\"ahler manifolds}
    The $n_V$ complex scalar fields of the gauge multiplet define a special K\"ahler manifold. Now we collect the needed facts about these manifolds. 
    
    Special K\"ahler manifolds are K\"ahler manifolds that possess a symplectic structure $Sp(2(n_V+1))$. 
    The basic ingredients are symplectic vectors 
    \begin{equation}
        V=\begin{pmatrix} X^I \\ F_I\end{pmatrix}=yv,\;\; v=\begin{pmatrix}Z^I(z)\\F_I(z)\end{pmatrix}_.
    \end{equation}
    The vector $V$ obeys the following
    \begin{equation}
        \langle V,\nabla_\alpha V\rangle=0,\;\;\;\nabla_\alpha V=y\nabla_\alpha v,\;\;\nabla_\alpha v=\partial_\alpha v+(\partial_\alpha \mathcal{K}) v.
    \end{equation}
    Here,
    \begin{equation}
        \langle V,\bar{V}\rangle \equiv X^I\bar{F_I}-F_I\bar{X^I}.
    \end{equation}
    The Kahler potential $\mathcal{K}$ is determined by 
    \begin{equation}
        \langle V,\bar{V}\rangle=i, (y\bar{y})^{-1}=e^{-\mathcal{K}}=-i\langle v,\bar{v}\rangle,
    \end{equation}
    which in turn defines the metric on the K\"ahler manifold as
    \begin{equation}
        g_{\alpha\bar{\beta}}=\partial_\alpha \partial_{\bar{\beta}}\mathcal{K}=i\langle \nabla_\alpha V, \bar{\nabla}_{\bar{\beta}}\bar{V}\rangle.
    \end{equation}
    The matrix ${\mathcal{N}}_{IJ}$, which couples the scalar fields to gauge fields, is defined as
    \begin{equation}
        \bar{\mathcal{N}}_{IJ}=(\bar{F}_I \;\;\;\nabla_\alpha F_I)(\bar{X}^J \;\;\;\nabla_\alpha X^J)^{-1}
    \end{equation}
    \subsection{Action}
    The action for this theory upto quadratic terms in fermions \footnote{we are studying fermion zero modes so the quartic fermionic terms, which would give cubic terms in fermions in the equation of motion, are irrelevant here} is \cite{freedman}
    \begin{eqnarray}
        \label{eq:fullAction}
        e^{-1}\mathcal{L}=\left(\frac{R}{2}-\bar{\psi}_{(i)\mu}\gamma^{\mu\nu\rho}D_\nu\psi_\rho^{(i)}\right)-g_{\alpha\bar{\beta}}\partial_\mu z^\alpha\partial^\mu\bar{z}^{\bar{\beta}}\nonumber\\
        +\Big(-\frac{1}{4}i\mathcal{N}_{IJ}F^{+I}_{\mu\nu}F^{+\mu\nu J}+F^{+I}_{\mu\nu}Im{\mathcal{N}}_{IJ}Q^{\mu\nu+J}\nonumber\\-\frac{1}{4}g_{\alpha\bar{\beta}}\bar{\chi}_{(i)}^\alpha \slashed{D}\chi^{(i)\bar{\beta}}+\frac{1}{2}g_{\alpha\bar{\beta}}\bar{\psi}_{(i)a}\slashed{\partial}z^\alpha\gamma^a\chi^{(i)\bar{\beta}}+h.c.\Big)
    \end{eqnarray}
    where 
    \begin{equation}
        Q^{ab-J}\equiv \bar{\nabla}_{\bar{\alpha}}\bar{X}^J\bar{\chi}^{\bar{\alpha}(i)}\gamma^a\psi^{b(j)}\varepsilon_{(ij)}+X^J\bar{\psi}^a_{(i)}\psi^b_{(j)}\varepsilon^{(ij)}+\frac{1}{8}\bar{\nabla}_{\bar{\alpha}}\bar{X}^J g^{\beta\bar{\alpha}}C_{\beta\gamma\delta}\bar{\chi}_{(i)}^\gamma \gamma^{ab}\chi_{(j)}^\delta\varepsilon^{(ij)}.
    \end{equation}
    The covariant derivatives are given by
    \begin{eqnarray}
        D_\mu\psi_{\nu (i)}&=&\left(\partial_\mu+\frac{1}{4}\omega_\mu^{ab}\gamma_{ab}+\frac{1}{2}i\mathcal{A}_\mu\right)\psi_{\nu (i)}\nonumber\\
        D_\mu \chi_{(i)}^\alpha&=&\left(\partial_\mu+\frac{1}{4}\omega_\mu^{ab}\gamma_{ab}+\frac{1}{2}i\mathcal{A}_\mu\right)\chi_{(i)}^\alpha+\Gamma^\alpha_{\beta\gamma}\chi_{(i)}^\gamma\partial_\mu z^\beta.
    \end{eqnarray}
    They contain the K\"ahler connection
    \begin{eqnarray}
        \mathcal{A}_\mu=\frac{1}{2}i\partial_\alpha \mathcal{K}\partial_\mu z^\alpha-\frac{1}{2}i\partial_{\bar{\alpha}}\mathcal{K}\partial_\mu\bar{z}^{\bar{\alpha}}.
    \end{eqnarray}
    \subsection{Fermion variation equations}
    The supersymmetry variation equations for the fermions are \cite{freedman}
    \begin{eqnarray}
    \label{eq:fermvar}
        \delta\psi^{(i)}_\mu&=&\left(\partial_\mu +\frac{1}{4}\omega^{ab}_\mu\gamma_{ab}-\frac{1}{2}i\mathcal{A}_\mu\right)\epsilon^{(i)}-\frac{1}{16}\gamma^{ab}T^-_{ab}\varepsilon^{(ij)}\gamma_\mu \epsilon_{(j)}\nonumber\\
        \delta \chi_{(i)}^\alpha&=&\slashed{\partial}z^\alpha \epsilon_{(i)}-\frac{1}{2}G^{-\alpha}_{ab}\gamma^{ab}\varepsilon_{(ij)}\epsilon^{(j)}
    \end{eqnarray}
    where,
    \begin{eqnarray}
        \label{eq:tgdef}
        T^+_{ab}=2i(\bar{X}^IG^+_{Iab}-\bar{F}_I F^{+I}_{ab})\nonumber\\
        G^{-\alpha}_{ab}=g^{\alpha\bar{\beta}}\bar{\nabla}_{\bar{\beta}}\bar{X}^I Im\,\mathcal{N}_{IJ}F^{-J}_{ab}.
    \end{eqnarray}
    \section{Denef Bates black holes}
    \label{sec:dbbh}
    A multicentered black hole solution for $\mathcal{N}=2$ Supergravity, described in \cite{denefbates}, is given by a BPS metric,
    \begin{equation}
        \label{eq:metric}
        ds^2=-e^{2U}(dt+\omega)^2+e^{-2U}dx^i dx^i
    \end{equation}
    where $U$ and $\omega$ together with the moduli fields $z^\alpha$, are time-independent solutions of the following equations
    \begin{eqnarray}
        2e^{-U}Im(e^{-i\alpha}V)=-H\nonumber\\
        *d\omega=\langle dH, H\rangle        \label{DBeq}
    \end{eqnarray}
    where $\alpha$ is an unknown real function, $H(x)$ is a given harmonic function, the Hodge star operates on flat $\mathbb{R}^3$.

    For $N$ charges located at $x_s$, $p=1,\cdots,N$; in asymptotically flat space, one has:
    \begin{equation}
        H=\sum_{s=1}^N\frac{\Gamma_s}{|x-x_s|}-2 Im(e^{-i\alpha}V)_{r=\infty}
    \end{equation}
    The boundary condition on $\alpha$ at $r=\infty$ is that it equals the phase of the total central charge, $\alpha=arg\,Z$. 
    \begin{equation}
        \label{eq:chargedef}
        \Gamma=\begin{pmatrix}
            p^I\\q_I
        \end{pmatrix}\;\;\; Z=2\langle V,\Gamma\rangle
    \end{equation}
    \ref{DBeq} gives us an integrability condition (by applying the operator $d*$ on both sides of the equation), which in turns gives us a restriction on the distances between the centers of the black holes
    \begin{equation}
        \langle\Delta H,H\rangle=0
    \end{equation}
    For a generic $H$, $\Delta H=\sum_{s=1}^N \Gamma_s\delta^3(x-x_s)$. Then the integrability conditon becomes $\langle \Gamma_s, H\rangle|_{x=x_s}=0$ or,
    \begin{equation}
        \sum_{r\neq s}\frac{\langle\Gamma_r,\Gamma_s\rangle}{|x_r-x_s|}=-2\langle \text{Im}(e^{-i\alpha}V)_{r=\infty},\Gamma_s\rangle
    \end{equation}
    this shows that the distances between the centres are constrained and the black holes form bound states. 
    
    This is an important ingredient, if suppose we are to study the effective supersymmetric quantum mechanics of the centres of the black holes then we would find that these constraints are constraints on the bosonic degrees of freedom, which would certainly reduce the fermionic degrees of freedom also. The fermion zero modes we are interested in here are these fermionic degrees of freedom in this effective theory.

    With no loss of generality we can assume $\alpha=0$, or $(Im\langle V,\Gamma\rangle)_{r=\infty}=0$. Then,
    \ref{DBeq} becomes 
    \begin{equation}
        ie^{-U}e^{\mathcal{K}/2}\begin{pmatrix}
         Z^I - \bar{Z}^I\\
         F_I -\bar{F}_I   
        \end{pmatrix}=H\equiv\begin{pmatrix}
            \tilde{H}^I\\
            H_I
        \end{pmatrix}
    \end{equation}
    or, 
    \begin{eqnarray}
        U=\frac{\mathcal{K}}{2}\nonumber\\
        i(v-\bar{v})=H
        \label{eq:metrickahler}.
    \end{eqnarray}
    Then by definition,
    \begin{eqnarray}
        e^{-2U}=e^{-\mathcal{K}}=-i(Z^I\bar{F}_I-F_I\bar{Z}^I)\nonumber\\
        \partial_\alpha \mathcal{K}=ie^{\mathcal{K}}(\partial_\alpha Z^I \bar{F}_I+Z^I \partial_\alpha \bar{F}_I-\partial_\alpha F_I\bar{Z}^I-F_I\partial_\alpha \bar{Z}^I)
    \end{eqnarray}
    since $F_I$ and $Z^I$ are holomorphic $\partial_\alpha\bar{F}_I=0$ and $\partial_\alpha \bar{Z}^I=0$.
    \begin{eqnarray}
        \partial_{{\alpha}}\mathcal{K}=ie^\mathcal{K}(\partial_\alpha Z^I\bar{F}_I-\partial_\alpha F_I \bar{Z}^I)\nonumber\\
        \partial_{\bar{\alpha}}\mathcal{K}=ie^\mathcal{K}(Z^I\partial_{\bar{\alpha}} \bar{F}_I-F_I\partial_{\bar{\alpha}}\bar{Z}^I)\nonumber\\
        \mathcal{A}_m=i\frac{1}{2}(\partial_\alpha \mathcal{K}\partial_m z^\alpha-\partial_{\bar{\alpha}}\mathcal{K}\partial_m \bar{z}^{\bar{\alpha}})\nonumber
    \end{eqnarray}
    \begin{eqnarray}
        \mathcal{A}_m=-\frac{1}{2}e^\mathcal{K}(\partial_m Z^I\bar{F}_I-\partial_m F_I \bar{Z}^I+F_I\partial_m \bar{Z}^I-Z^I\partial_m\bar{F}_I)
    \end{eqnarray}
    also from $\langle v,\nabla_\alpha v\rangle=Z^I\partial_\alpha F_I-F_I\partial_\alpha Z^I=0.$ we have
    \begin{eqnarray}
        \mathcal{A}_m&=&-\frac{1}{2}e^\mathcal{K}\left[(Z^I-\bar{Z}^I)\partial_m(F_I-\bar{F}_I)-(F_I-\bar{F}_I)\partial_m(Z^I-\bar{Z}^I)\right]\nonumber\\
        \mathcal{A}&=&\frac{1}{2}e^{2U}\langle{d(v-\bar{v})},(v-\bar{v})\rangle\nonumber\\
        \mathcal{A}&=&\frac{1}{2}e^{2U}\langle{H,dH}\rangle\nonumber\\
        \mathcal{A}&=&-\frac{1}{2}e^{2U}*d\omega
        \label{eq:connA}
    \end{eqnarray}
    \ref{eq:metrickahler} and \ref{eq:connA} are the same equations which define black holes in \cite{BLS}.

    The gauge field of the background is given by \cite{denefbates}
    \begin{eqnarray}
        F^I_{ij}=e^{2U}\epsilon_{ijk}\delta^k_m\partial_m\tilde{H}^I\nonumber\\
        G_{Iij}=e^{2U}\epsilon_{ijk}\delta^k_m\partial_m H_I.
    \end{eqnarray}
    where,
    \begin{eqnarray}
        G^+_{\mu\nu I}=\mathcal{N}_{IJ}F^{+J}_{\mu\nu}.
    \end{eqnarray}

    \subsection{Killing Spinors}
    These black holes are supersymmetric by construction. We can also check their supersymmetry by solving the fermion variation equations \ref{eq:fermvar} to find the Killing spinors. The Killing spinors are needed to find the broken zero modes. 
    
    The vierbeins for the Denef Bates black hole are
    \begin{equation}
        e^0=e^U(dt+\omega),\;\; e^i=e^{-U}\delta^i_m \;dx^m.
    \end{equation}
    In appendix \ref{app:spinc} we show that the spin connections for this background are,
    \begin{eqnarray}
        \omega^{-0i}_t&=&\frac{1}{2}e^{2U}\delta^i_m(\partial_m U+i\mathcal{A}_m)=-\frac{1}{4}e^{U}T^{-0i}\nonumber\\
        \omega^{-ij}_t&=&\frac{1}{2}ie^{2U}\epsilon_{ijk}\delta^k_m(\partial_m U+i\mathcal{A}_m)=\frac{1}{4}ie^{U}\epsilon_{ijk}T^-_{0k}=-\frac{1}{4}e^UT^{-ij}\nonumber\\
        \omega_m^{-0i}&=&-\frac{1}{4}e^U\omega_mT^{-0i}+\frac{i}{4}e^{-U}\delta^i_n\epsilon_{nmk}T^{-0k}.
    \end{eqnarray}
    Computing the time component of the gravitini variation equation \ref{eq:fermvar}
    \begin{eqnarray}
        \delta\psi^{(i)}_t=\left(\partial_t +\frac{1}{4}\omega^{ab}_t\gamma_{ab}-\frac{1}{2}i\mathcal{A}_t\right)\epsilon^{(i)}-\frac{1}{16}\gamma^{ab}T^-_{ab}\varepsilon^{(ij)}\gamma_t \epsilon_{(j)}\nonumber
    \end{eqnarray}
    \begin{equation}
        \partial_t\epsilon^{(i)}=0\;\;\;\mathcal{A}_t=0\;\;\; \gamma_t=e^U\gamma_0
    \end{equation}
    \begin{eqnarray}
        \delta\psi^{(i)}_t= -\frac{1}{16}e^UT^{-}_{ab}\gamma^{ab}\epsilon^{(i)}-\frac{1}{16}e^U\gamma^{ab}T^-_{ab}\varepsilon^{(ij)}\gamma_0 \epsilon_{(j)}=-\frac{1}{16}e^UT^{-}_{ab}\gamma^{ab}(\epsilon^{(i)}+\gamma_0\varepsilon^{(ij)}\epsilon_{(j)})\nonumber
    \end{eqnarray}
    This shows that the Killing spinors obey,
    \begin{eqnarray}
        \epsilon^{(i)}=-\gamma_0\varepsilon^{(ij)}\epsilon_{(j)}\nonumber\\
        \gamma_0\epsilon^{(i)}=\varepsilon^{(ij)}\epsilon_{(j)}
    \end{eqnarray}
    Now for the spatial components
    \begin{eqnarray}
        \delta\psi^{(i)}_m=\left(\partial_m +\frac{1}{4}\omega^{ab}_m\gamma_{ab}-\frac{1}{2}i\mathcal{A}_m\right)\epsilon^{(i)}-\frac{1}{16}\gamma^{ab}T^-_{ab}\varepsilon^{(ij)}\gamma_m \epsilon_{(j)}\nonumber
    \end{eqnarray}
    \begin{equation}
        \gamma_m=e^U\omega_m\gamma_0+e^{-U}\delta^k_m\gamma_k\nonumber
    \end{equation}
    \begin{eqnarray}
    \delta\psi^{(i)}_m=\left(\partial_m-\frac{1}{2}i\mathcal{A}_m\right)\epsilon^{(i)}-\left(\frac{1}{4}e^U\omega_m T^{-0i}\gamma_{0i}P_L-\frac{i}{4}e^{-U}\epsilon_{ijk}\delta^j_mT^{-0k}\gamma_{0i}P_L\right)\epsilon^{(i)}\nonumber\\
    -\frac{1}{4}T^{-0i}\gamma_{0i}P_L(e^U\omega_m\gamma_0+e^{-U}\delta^k_m\gamma_k)\varepsilon^{(ij)}\epsilon_{(j)}\nonumber\\
    =\left(\partial_m-\frac{1}{2}i\mathcal{A}_m\right)\epsilon^{(i)}-\frac{1}{4}e^U\omega_mT^{-0i}\gamma_{0i}(\epsilon^{(i)}+\gamma_0\varepsilon^{(ij)}\epsilon_{(j)})\nonumber\\
    -\frac{1}{4}e^{-U}\left(-i\epsilon_{ijk}\delta^j_mT^{-0k}\gamma_{0i}\epsilon^{(i)}+T^{-0i}\gamma_{0i}\delta^k_m\gamma_k\varepsilon^{(ij)}\epsilon_{(j)}\right)\nonumber
    \end{eqnarray}
    \begin{equation}
        i\epsilon_{ijk}\gamma_{0i}\epsilon^{(i)}=\gamma_{jk}\epsilon^{(i)}
    \end{equation}
    for the Killing spinor, $\epsilon^i=-\gamma_0\varepsilon^{ij}\epsilon_j$,
    \begin{eqnarray}
        \delta\psi^{(i)}_m=\left(\partial_m-\frac{1}{2}i\mathcal{A}_m\right)\epsilon^{(i)}+\frac{1}{4}e^{-U}\delta^j_mT^{-0j}\epsilon^{(i)}\nonumber\\
        \delta\psi^{(i)}_m=\left(\partial_m-\frac{1}{2}i\mathcal{A}_m\right)\epsilon^{(i)}-\frac{1}{2}(\partial_m U+i\mathcal{A}_m)\epsilon^{(i)}\nonumber\\
        \delta\psi^{(i)}_m=\left(\partial_m-\frac{1}{2}\partial_m U-i\mathcal{A}_m\right)\epsilon^{(i)}
    \end{eqnarray}
    Hence the Killing spinor is a solution of the equation
    \begin{eqnarray}
        \left(\partial_m-\frac{1}{2}\partial_m U-i\mathcal{A}_m\right)\epsilon^{(i)}=0.
    \end{eqnarray}
    The integrability condition for this equation is 
    \begin{equation}
        \label{eq:da0}
        d\mathcal{A}=0.
    \end{equation}
    The Gaugino variation equation is given by \ref{eq:fermvar}
    \begin{eqnarray}
        \delta \chi_i^\alpha=\slashed{\partial} z^\alpha \epsilon_{(i)}-\frac{1}{2}G^{-\alpha}_{ab}\gamma^{ab}\varepsilon_{(ij)}\epsilon^{(j)}\nonumber
    \end{eqnarray}
    \begin{equation}
        G^{-\alpha}_{ab}=g^{\alpha\bar{\beta}}\bar{\nabla}_{\bar{\beta}}\bar{X}^I Im\;\mathcal{N}_{IJ}F^{-J}_{ab}\nonumber
    \end{equation}
    from appendix \ref{app:spinc} 
    \begin{equation}
        G^{-\beta}_{0i}=\frac{1}{2}e^U\delta^i_m\partial_m z^\beta
    \end{equation}
    substituting in the gaugini variation equation
    \begin{eqnarray}
        \delta\chi_{(i)}^\alpha=\slashed{\partial}z^\alpha\epsilon_{(i)}-2G^{-\alpha}_{0i}\gamma^{0i}\varepsilon_{(ij)}\epsilon^{(j)}\nonumber\\
        =\slashed{\partial}z^\alpha\epsilon_{(i)}-e^U\partial_m z^\beta\gamma^i\delta^i_m\epsilon_{(i)}=0
    \end{eqnarray}
    The gaugini variation equations do not impose any more constraints. Hence, we have shown that the Denef Bates solutions are supersymmetric and explicitly found the Killing spinors.
    \section{Fermion zero modes}
    \label{sec:fzms}
    The Killing spinors are defined by
    \begin{eqnarray}
        &&\gamma_0\epsilon^{(i)}=\varepsilon^{(ij)}\epsilon_{(j)}\nonumber\\
        &&\left(\partial_m-\frac{1}{2}\partial_m U-i\mathcal{A}_m\right)\epsilon^{(i)}=0.
    \end{eqnarray}
    The orthogonal space of spinors which breaks supersymmetry for the Denef Bates background is defined by
    \begin{eqnarray}
        \label{eq:antiKill}
        &&\gamma_0\epsilon^{(i)}=-\varepsilon^{(ij)}\epsilon_{(j)}\nonumber\\
        &&\left(\partial_m-\frac{1}{2}\partial_m U-i\mathcal{A}_m\right)\epsilon^{(i)}=0.
    \end{eqnarray}
    This space is called the anti Killing spinor space, and since these spinors break supersymmetry
    we can use these spinors to generate extra solutions to the equations of motion. Here we are
    interested only in counting fermion zero modes hence we are not concerned with the back reaction of these modes.

    The fermion zero modes can be found by using our anti Killing spinors defined by \ref{eq:antiKill}
    in the supersymmetry variation equations \ref{eq:fermvar}. Doing so we find,
    \begin{eqnarray}
        \label{eq:fzm}
        \psi_0^{(i)}=-\frac{1}{2}T^{-0i}\gamma_{0i}\epsilon^{(i)}\nonumber\\
        \psi_j^{(i)}=-\frac{1}{2}T^{-0i}\gamma_{i}\gamma_{j}\epsilon^{(i)}\nonumber\\
        \chi_{(i)}^\alpha=2\slashed{\partial}z^\alpha\epsilon_{(i)}.
    \end{eqnarray}
    If these are fermion zero modes they will satisfy the equations of motion for the fermions. We will now check this.
    \subsection{Equations of motion for the fermions}
        The equations of motion for the fermions are shown in appendix \ref{app:eomf} to be
        \begin{eqnarray}
            -\gamma^{abc}D_b\psi_c^{(i)}-g_{\alpha\bar{\beta}}G^{+\bar{\beta}ab}\gamma_b\chi^\alpha_j\varepsilon^{(ij)}-\frac{1}{2}T^{-ab}(\psi_{(j)})_b\varepsilon^{(ij)}+\frac{1}{2}g_{\alpha\bar{\beta}}\slashed{\partial}z^\alpha\gamma^a\chi^{(i)\bar{\beta}}=0\nonumber\\
            \frac{1}{4}G^{+\bar{\gamma}ab}C_{\bar{\gamma}\bar{\beta}\bar{\delta}}\gamma_{ab}\chi^{(j)\bar{\delta}}\varepsilon_{(ij)}
        +g_{\alpha\bar{\beta}}G^{-\alpha ab}\gamma_a\psi^{(j)}_b\varepsilon_{(ij)}-\frac{1}{2}g_{\alpha\bar{\beta}}\slashed{D}\chi_{(i)}^{\alpha}=0.
        \label{eq:eom}
        \end{eqnarray}
        For $a=0$ the gravitini equation of motion is,
        \begin{eqnarray}
            -\gamma^{0ij}D_i\psi_j^{(i)}-g_{\alpha\bar{\beta}}G^{+\bar{\beta}0i}\gamma_i\chi_{(j)}^\alpha \varepsilon^{(ij)}-\frac{1}{2}T^{-0i}\gamma_{i(j)}\varepsilon^{(ij)}+\frac{1}{2}g_{\alpha\bar{\beta}}\slashed{\partial}z^\alpha\gamma^0\chi^{(i)\bar{\beta}}=0
        \end{eqnarray}
        We can show that,
        \begin{eqnarray}
            \gamma_0\gamma_{ij}D_i\psi_j^{(i)}&=&-2e^{2U}\left(\partial_m(\partial_m U+i\mathcal{A}_m)\right)\gamma_0\epsilon^{(i)}\nonumber\\
        &&-e^{2U}\left(\partial_m U-i\mathcal{A}_m\right)\left(\partial_m U+i\mathcal{A}_m\right)\gamma_0\epsilon^{(i)}\nonumber\\
        &&+2e^{2U}\left(\partial_m U+i\mathcal{A}_m\right)\left(\partial_m U+i\mathcal{A}_m\right)\gamma_0\epsilon^{(i)}.
        \end{eqnarray}
        \begin{eqnarray}
        -\frac{1}{2}T^{-0j}\psi_{(j)j}\varepsilon^{(ij)}=e^{2U}(\partial_m U+i\mathcal{A}_m)\delta^j_m(\partial_n U-i\mathcal{A}_n)\delta^k_n\gamma_k\gamma_j\epsilon_{(j)}\varepsilon^{(ij)}\nonumber\\
        =e^{2U}(\partial_m U\partial_m U+\mathcal{A}_m\mathcal{A}_m)\epsilon_{(j)}\varepsilon^{(ij)}+ie^{2U}(\partial_n U\mathcal{A}_m-\partial_m U\mathcal{A}_n)\delta^j_m\delta^k_n\gamma_{kj}\epsilon_{(j)}\varepsilon^{(ij)}\nonumber\\
    \end{eqnarray}
    \ref{eq:da0} implies $\partial_n U\mathcal{A}_m-\partial_m U\mathcal{A}_n=0$
    \begin{eqnarray}
        \gamma_0\gamma_{ij}D_i\psi_j^{(i)}-\frac{1}{2}T^{-0j}\psi_{(j)j}\varepsilon^{(ij)}=-2e^{2U}\left(\partial_m(\partial_m U+i\mathcal{A}_m)\right)\varepsilon^{(ij)}\epsilon_{(j)}\nonumber\\
        +2e^{2U}\left(\partial_m U+i\mathcal{A}_m\right)\left(\partial_m U+i\mathcal{A}_m\right)\varepsilon^{(ij)}\epsilon_{(j)}
    \end{eqnarray}
    \begin{eqnarray}
        -g_{\alpha\bar{\beta}}G^{+\bar{\beta}0i}\gamma_i\chi^\alpha_j\varepsilon^{ij}+\frac{1}{2}g_{\alpha\bar{\beta}}\slashed{\partial}z^\alpha\gamma^0\chi^{i\bar{\beta}}=2e^{2U}g_{\alpha\bar{\beta}}\partial_m z^\alpha\partial_m \bar{z}^{\bar{\beta}}\varepsilon^{(ij)}\epsilon_{(j)}\nonumber
    \end{eqnarray}
    \begin{eqnarray}
        \partial_m(\partial_m U+i\mathcal{A}_m)=g_{\alpha\bar{\beta}}\partial_m z^\alpha\partial_m z^{\bar{\beta}}+\partial_{\bar{\alpha}}\mathcal{K}\partial_{\bar{\beta}}\mathcal{K}\partial_m \bar{z}^{\bar{\alpha}}\partial_m \bar{z}^{\bar{\beta}}
    \end{eqnarray}
    Hence, we can see that the fermion zero modes satisfy the $a=0$ equation of motion,
    \begin{eqnarray}
         \gamma_0\gamma_{ij}D_i\psi_j^{(i)}-\frac{1}{2}T^{-0j}\psi_{(j)j}\varepsilon^{(ij)}-g_{\alpha\bar{\beta}}G^{+\bar{\beta}0i}\gamma_i\chi^\alpha_j\varepsilon^{ij}+\frac{1}{2}g_{\alpha\bar{\beta}}\slashed{\partial}z^\alpha\gamma^0\chi^{i\bar{\beta}}=0\nonumber.
    \end{eqnarray}
    By similar substitutions, we can show that the equation of motion for $a=i$ and also the gaugini equations of motion \ref{eq:eom} are satisfied.

    \subsection{Supercharge, boundary conditions and norm}
    The supercharge for this theory has been shown to be \cite{supercharge,Teitelboim}
    \begin{eqnarray}
        Q^{(i)}=-i\int_{S^2_\infty}d\Sigma_i\gamma^{0ij}\psi_j^{(i)}.
    \end{eqnarray}
    Using \ref{eq:fzm} and \ref{eq:antiKill} we get
    \begin{eqnarray}
        Q^{(i)}=-2i\int_{S^2_\infty}d\Sigma_i T^{-0i}\varepsilon^{(ij)}\epsilon_{(j)}\nonumber
    \end{eqnarray}
    using \ref{eq:tgdef} and \ref{eq:chargedef}
    \begin{eqnarray}
        Q^{(i)}=2Z|_{r=\infty}\varepsilon^{(ij)}\epsilon_{(j)}.
    \end{eqnarray}
    To have a sensical supercharge the fermion zero modes must behave asymptotically as
    \begin{eqnarray}
        r^2\psi_\mu\xrightarrow{r=\infty}c_\mu,
    \end{eqnarray}
    where $c_\mu$ is a constant spinor.
   
    The norm for fermion zero modes was introduced in \cite{kallosh}, it was used by the authors to show that extremal Reissner-Nordstr\"om and extremal Dilation black holes have normailizable fermion zero modes. We borrow the same norm here,
    \begin{eqnarray}
        \label{eq:norm}
        ||\psi||^2 =\int_\Sigma d^3x \sqrt{-g_{(3)}}\psi_{\mu{(i)}}^\dagger\psi_{\nu}^{(i)}g^{\mu\nu}=\int_\Sigma d\Sigma_\mu\bar{\psi}_{\nu(i)}\gamma^{\mu\nu\rho}\psi_{\rho}^{(i)}
    \end{eqnarray}
    where $\Sigma$ is a space-like hypersurface and $g_{(3)}$ is the determinant of the metric induced on it. We take $\Sigma$ to be $\mathbb{R}^3$ and substitute \ref{eq:fzm}
    \begin{eqnarray}
        ||\psi||^2=\frac{1}{4}||\epsilon||^2\int_{\mathbb{R}^3}d^3 x \;((\partial_m U)^2+(\mathcal{A}_m)^2)
    \end{eqnarray}
    Hence, generally we have $||\psi||^2>0$. The goldstino modes are normalizable. 
    Now, we can search for extra fermion zero modes which are normalizable and have the same boundary conditions as the goldstino modes.
    \section{Extra fermion zero modes}
    \label{sec:xfzm}
    If extra fermion zero modes exist then they must be necessarily gauge inequivalent to the goldstino modes we have constructed in \ref{sec:fzms}. Now, we ask the question when is it that two configurations $\psi$ and $\xi$ have the same supercharge but are not gauge equivalent
    \begin{eqnarray}
        Q^{(i)}=-i\int_{S^2_\infty}d\Sigma_i\gamma^{0ij}\psi_j^{(i)}=-i\int_{S^2_\infty}d\Sigma_i\gamma^{0ij}\xi_j^{(i)}
    \end{eqnarray}
    This happens if and only if there exists a non trivial zero charge gravitini, which is not gauge equivalent to zero. A zero charge gravitini has the boundary condition
    \begin{eqnarray}
        r^2\psi\xrightarrow{r=\infty}0.
    \end{eqnarray}
    We should  note that the goldstino gravitini \ref{eq:fzm} are continious all over $\mathbb{R}^3$, which can be seen from the continuity of $T^{-0i}$ at the centers
    \begin{eqnarray}
        T^{-0i}=2e^{3U}\delta^i_m(Z^I\partial_m H_I-F_I\partial_m \tilde{H}^I)\nonumber\\
        e^{-U}\sim r^{-1},\;\; \partial_m H_I\sim \frac{x_m}{r^3}e^{3U},\;\; T^{-0i}\sim \delta^i_m \frac{x_m}{r}.
    \end{eqnarray}
    Also, since in a normed vector space every non zero state is normalizable, we demand that these zero charge modes are normalizable.
    \subsection{Single center}
    We first consider the single centered Denef Bates black hole. 
    For such backgrounds $\omega$ in \ref{eq:metric} is zero, 
    \begin{eqnarray}
        ds^2=-e^{2U}dt^2+e^{-2U}(dr^2+r^2d\Omega^2)
    \end{eqnarray}
    here we are looking for fermion zero modes which are solutions to the equation of motion, satisfy the boundary conditions $r^2\psi_\mu\xrightarrow{r=\infty}0$, continuity of the zero modes, and are normalizable.

    Normalizability implies that the integral 
    \begin{equation}
        \int_\Sigma d^3x \sqrt{-g_{(3)}}\psi_{\mu{(i)}}^\dagger\psi_{\nu}^{(i)}g^{\mu\nu}
    \end{equation}
    is finite. The asymptotic boundary condition implies that the integral will only recieve contributions from  the near horizon limit. 
    \begin{equation}
        \int_\Sigma dr\,d\theta\,d\phi e^{-3U}r^2\sin\theta \left(e^{-2U}\psi_{t(i)}^\dagger\psi_t^{(i)} +e^{2U}\psi_{r(i)}^\dagger\psi_r^{(i)}+r^{-2}e^{2U}\psi_{\theta(i)}^\dagger\psi_\theta^{(i)}+r^{-2}e^{2U}sin^{-2}\theta\psi_{\phi(i)}\psi_\phi^{(i)}\right)
    \end{equation}
    $e^U$ approches 1 asymptotically. This integral recieves no contribution in the asymptotic limit.

    Hence, we need to study the solutions of the equation of motion in the near horizon limit with the fermion zero mode continuous at $r=0$.

    The most general background for a single centre black hole in the near horizon limit is
    \begin{eqnarray}
        &&ds^2=a^2\left(-r^2\, dt^2+\frac{dr^2}{r^2}\right)+a^2\left(d\theta^2+\sin^2\theta d\phi^2\right),\nonumber\\
        &&z^\alpha=u^\alpha,\;\; F^I_{tr}=q^I \label{eq:nhmet}
    \end{eqnarray}
    where $a$, $u^\alpha$ and $q^I$ are constants which can be determined by Sen's entropy function formalism. This is a maximally supersymmetric background.

    The norm in this limit is
    \begin{eqnarray}
        ||\psi||^2=\int_\Sigma dr\,d\theta\,d\phi \frac{a^3\sin\theta}{r}\left(\frac{1}{a^2r^2}\psi^{\dagger}_{(i)t}\psi^{(i)}_t+\frac{r^2}{a^2}\psi_{r(i)}^\dagger\psi_r^{(i)}+\frac{1}{a^2}\psi_{\theta(i)}^\dagger\psi_\theta^{(i)}+\frac{1}{a^2\sin^2\theta}\psi^\dagger_{\phi(i)}\psi_\phi^{(i)}\right)
    \end{eqnarray}
    
    This shows that to have a finite non zero contribution in the near horizon limit to the norm we must have 
    \begin{eqnarray}\label{eq:contribnorm}
        \psi_{t(i)}&\sim& r^n\;\;n\in[0,3/2]\nonumber\\
        \psi_{r(i)}&\sim &r^0\nonumber\\
        \psi_{\theta(i)}&\sim &r^n\;\;n\in[0,1/2]\nonumber\\
        \psi_{\phi(i)}&\sim& r^n\;\;n\in[0,1/2]
    \end{eqnarray}
    the lower limit on $n$ is fixed by requiring the gravitini to be continious all over $\mathbb{R}^3$. 
    \subsection{Near horizon equation of motion}
    We transform the metric to
    \begin{eqnarray}
        ds^2=\left(-\frac{r^2}{a^2}{dt^2}+a^2\frac{dr^2}{r^2}\right)+a^2(d\theta^2+\sin^2\theta\,d\phi^2)
    \end{eqnarray}
    This shows 
    \begin{eqnarray}
        e^{U}=\frac{r}{a},\;\;\; \partial_r U=\frac{1}{r},\;\; \partial_\theta U=0,\;\;\;\partial_\phi U=0\nonumber\\
    \end{eqnarray}
    Since, the scalars are constants all over space in the near horizon limit $\partial_\mu z^\alpha=0$ ($G^{-\alpha}_{ab}=0$). This reduces the terms in the equation of motion for the gravitini \ref{eq:eom}
    \begin{eqnarray}
        \gamma^{abc}D_b\psi_c^{(i)}+\frac{1}{2}T^{-ab}\psi_{b(j)}\varepsilon^{(ij)}=0
    \end{eqnarray}
    The spin connections are
    \begin{eqnarray}
        \omega^{01}_t=\frac{r}{a^2},\;\;\omega^{23}_\phi=-\cos\theta,
    \end{eqnarray}
    and the tensor T is given by
    \begin{eqnarray}
        T^{-\theta\phi}=-\frac{2i}{a^3\sin\theta},\;\;T^{-tr}=-\frac{2}{a}.
    \end{eqnarray}
    The equations of motion then are
    \begin{eqnarray}
        &&\gamma^{012}\frac{1}{a}\left(\partial_r\psi_\theta^{(i)}-\partial_\theta\psi_r^{(i)}\right)+\gamma^{013}\frac{1}{a\sin\theta}\left(\partial_r\psi_\phi^{(i)}-\partial_\phi \psi_r^{(i)}+\frac{1}{2}\cos\theta\;\gamma_{23}\;\psi_r^{(i)}\right)\nonumber\\
        &&+\gamma^{023}\frac{1}{ar\sin\theta}\left(\partial_\theta\psi_\phi^{(i)}-\partial_\phi\psi_\theta^{(i)}+\frac{1}{2}\cos\theta\,\gamma_{23}\,\psi_\theta^{(i)}\right)-\frac{1}{a}\psi_{r(j)}\varepsilon^{(ij)}=0\nonumber\\
        &&\gamma^{102}\frac{1}{a}\left(\frac{1}{2}\frac{r}{a^2}\gamma_{01}\psi_\theta^{(i)}-\partial_\theta\psi_t^{(i)}\right)+\frac{1}{a\sin\theta}\gamma^{103}\left(\frac{1}{2}\frac{r}{a^2}\gamma_{01}\psi_\phi^{(i)}-\partial_\phi\psi_t^{(i)}+\frac{1}{2}\cos\theta\,\gamma_{23}\,\psi_{t}^{(i)}\right)\nonumber\\
        &&+\gamma^{123}\frac{r}{a^3\sin\theta}\left(\partial_\theta\psi_\phi^{(i)}-\partial_\phi\psi_\theta^{(i)}+\frac{1}{2}\cos\theta\,\gamma_{23}\,\psi_\theta^{(i)}\right)+\frac{1}{a}\psi_{t(j)}\varepsilon^{(ij)}=0\nonumber\\
        &&\gamma^{201}\frac{1}{a}\left(\frac{1}{2}\frac{r}{a^2}\gamma_{01}\psi_r^{(i)}-\partial_r\psi_t^{(i)}\right)+\gamma^{203}\frac{1}{ar\sin\theta}\left(\frac{1}{2}\frac{r}{a^2}\gamma_{01}\psi_\phi^{(i)}-\partial_\phi\psi_t^{(i)}+\frac{1}{2}\cos\theta\,\gamma_{23}\,\psi_{t}^{(i)}\right)\nonumber\\
        &&+\gamma^{213}\frac{r}{a^3 \sin\theta}\left(\partial_r\psi_\phi^{(i)}-\partial_\phi \psi_r^{(i)}+\frac{1}{2}\cos\theta\;\gamma_{23}\;\psi_r^{(i)}\right)-\frac{i}{a^3\sin\theta}\psi_{\phi(j)}\varepsilon^{(ij)}=0\nonumber\\
        &&\gamma^{301}\frac{1}{a\sin\theta}\left(\frac{1}{2}\frac{r}{a^2}\gamma_{01}\psi_r^{(i)}-\partial_r\psi_t^{(i)}\right)+\gamma^{302}\frac{1}{ar\sin\theta}\left(\frac{1}{2}\frac{r}{a^2}\gamma_{01}\psi_\theta^{(i)}-\partial_\theta\psi_t^{(i)}\right)\nonumber\\
        &&+\gamma^{312}\frac{r}{a^3\sin\theta}\left(\partial_r\psi_\theta^{(i)}-\partial_\theta\psi_r^{(i)}\right)+\frac{i}{a^3\sin\theta}\psi_{\theta(j)}\varepsilon^{(ij)}=0
    \end{eqnarray}
    If we assume that $\psi_t^{(i)}\sim r^a$, $\psi_r^{(i)}\sim r^b$, $\psi_\theta^{(i)}\sim r^c$, $\psi_\phi\sim r^d$. Then from the above equations we get the relations
    \begin{eqnarray}
        b=c-1=d-1\nonumber\\
        a=c+1=d+1\nonumber\\
        b+1=a-1=d\nonumber\\
        b+1=a-1=c\nonumber
    \end{eqnarray}
    which shows that 
    \begin{eqnarray}
        b=a-2=c-1=d-1.\label{eq:rdepend}
    \end{eqnarray}
    The lowest value of $b$ that can occur is $0$ (due to continuity). Which shows that the lowest values of these constants are $a=2, b=0, c=1, d=1$. For these values from \ref{eq:contribnorm} we get no contribution to the norm. 

    The gauge condition $\gamma^a\psi^{(i)}_a=0$ shows
    \begin{eqnarray}
        \frac{a}{r}\gamma^0\psi^{(i)}_t+\frac{r}{a}\gamma^1\psi^{(i)}_r+\frac{1}{a}\gamma^2\psi^{(i)}_\theta+\frac{1}{a\sin\theta}\gamma^3\psi^{(i)}_\phi=0.
    \end{eqnarray}
    This also agrees with \ref{eq:rdepend} .

    Thus the dependence of $r$ in the equation of motion and the gauge condition shows that to have a contribution to the norm we must have a fermion zero mode which is discontinious at $r=0$; which would disagree with the boundary conditions of the broken modes.
    \subsection{Multicentred black holes}
    For multicentred Denef Bates black holes the near horizon metric for any of the horizons is \ref{eq:nhmet} and the same arguments will show that near every horizon there is no contribution to the norm. And, for zero charge black holes we must also have no contribution to the norm in the asymptotic limit. Thus the norm would vanish, and consecutively the black holes will contribute to the index. 
    \section{Conclusion}
    We set out to show that there are no extra fermion zero modes for multicentred Denef Bates black holes apart from the goldstino zero modes.
    We constructed the goldstino modes and computed their norm and supercharge.
    To find solutions to the fermion equations of motion, we argued that if such solutions exist  they must be exist as normalizable zero charge solutions.
    From the boundary condition of the zero charge solutions we showed that the norm will recieve zero contribution in the asymptotic limit.
    Then the only contribution to the norm can come from the near horizon limit.
    We wrote our norm in the near horizon limit and we saw that for small $r$ if there is a contribution to then the gravitini will be discontinious. But, our fermion zero modes are required to be continious all over $\mathbb{R}^3$.
    This shows that the norm of such zero charge modes must be zero.
    Hence, as argued in the introduction we have proved that multicentered Denef Bates black holes contribute to the index.

    The goldstino fermion zero modes are hair degrees of freedom of the black hole since they have support at infinity. The charge and the norm of these modes depend directly on the value of the fermion zero mode at infinity. In trying to solve this problem we showed that to have gauge inequivalent extra modes we must construct fermion zero modes which have zero charge which are not gauge equivalent to zero. Since they have zero charge they must be zero at infinity. Hence these modes we set out to find are not hair modes. If these modes were present they would have some corrections at the horizon of the black hole.

    In \cite{Banerjee:2008pv} the authors have argued that in $\mathcal{N}=4$ Supergravity there exist dyons which can decay into multicentered bound states. There they have assumed the result that such decays do not contribute to the index due to presence of additional fermion zero modes. This work has been a prelude to checking this more complicated setting. Also, constructing extra zero modes can be used to determine if the fuzzball configurations contribute to the index. We hope to report progress on these fronts in a future paper.

    \appendix
    \section{Spin connection, T and G}
    \label{app:spinc}
    The vierbeins are
    \begin{equation}
        e^0=e^U(dt+\omega),\;\; e^i=e^{-U}\delta^i_m \;dx^m
    \end{equation}
    finding spin connections
    \begin{eqnarray}
        de^0=e^U\,dU\wedge (dt+\omega)+e^U d\omega\;\nonumber\\
        de^i=-e^{-U}\,dU\wedge dx^m\delta^i_m
    \end{eqnarray}
    substituting in the structure equation $de^a+\omega^a_{\;b}\wedge e^b=0$
    \begin{equation}
        de^0+\omega^{0i}\wedge e^i=0,\;\; de^i+\omega^{ij}\wedge e^j+\omega^{0i}\wedge e^0=0
    \end{equation}
    Expanding the first equation we get
    \begin{eqnarray}
        de^0+e^{-U}(\omega^{0i})_t dt\wedge\delta^i_m dx^m+e^{-U}(\omega^{0i})_n dx^n\wedge\delta^i_m dx^m=0\nonumber\\
        e^U\,dU\wedge dt+e^{-U}(\omega^{0i})_t dt\wedge\delta^i_m dx^m=0\nonumber\\
        e^U\,dU\wedge\omega+e^U d\omega+e^{-U}(\omega^{0i})_n dx^n\wedge\delta^i_m dx^m=0\nonumber\\
        e^U\,\partial_m U=e^{-U}(\omega^{0i})_t\delta^i_m\nonumber\\
        e^U dU\wedge\omega+e^U\,d\omega+e^{-U}(\omega^{0i})_n dx^n\wedge\delta^i_m dx^m=0
    \end{eqnarray}
    define $\omega_{nm}=\frac{1}{2}(\partial_n\omega_m-\partial_m\omega_n)$
    \begin{eqnarray}
        e^{2U}\partial_n U\omega_m+e^{2U}\omega_{nm}+(\omega^{0i})_n\delta^i_m=0
    \end{eqnarray}
    Expanding the second equation we get
    \begin{eqnarray}
        de^i+\omega^{ij}\wedge e^j+\omega^{0i}\wedge e^0=0\nonumber\\
        de^i+e^{-U}\delta^j_m(\omega^{ij})_t dt\wedge dx^m+e^{-U}\delta^j_m(\omega^{ij})_n dx^n\wedge dx^m\nonumber\\
        +e^U(\omega^{0i})_m\,dx^m\wedge (dt+\omega)+e^U(\omega^{0i})_t dt\wedge\omega=0\nonumber\\
        e^{-U}\delta^j_m(\omega^{ij})_t dt\wedge dx^m+e^U(\omega^{0i})_m dx^m\wedge dt+e^U(\omega^{0i})_t\omega_mdt\wedge dx^m=0\nonumber\\
        \delta^j_m(\omega^{ij})_t=e^{2U}(\omega^{0i})_m-e^{2U}(\omega^{0i})_t\omega_m\nonumber\\
        \delta^j_m(\omega^{ij})_t=e^{4U}\delta^n_i(\partial_n U\omega_m+\omega_{nm})-e^{4U}\delta^n_i\partial_n U \omega_m\nonumber\\
        (\omega^{ij})_t=e^{4U}\delta^i_n\delta^j_m\omega_{nm}
    \end{eqnarray}
    \begin{eqnarray}
        de^i+e^{-U}\delta^j_m(\omega^{ij})_n dx^n\wedge dx^m+e^U(\omega^{0i})_m \omega_n dx^m\wedge dx^n=0\nonumber\\
        \delta^j_m(\omega^{ij})_n dx^n\wedge dx^m=\partial_n U dx^n\wedge dx^m\delta^i_m+e^{4U}\delta^i_p(\partial_p U\omega_m+\omega_{pm})\omega_n\, dx^m\wedge dx^n\nonumber\\
        \delta^j_m(\omega^{ij})_n =-\partial_{[m} U \delta^i_{n]}-e^{4U}\delta^i_p\omega_{p[m}\omega_{n]}\nonumber\\
    \end{eqnarray}
    We also have from \ref{eq:connA},
    \begin{equation}
        \omega_{nm}=-e^{-2U}\epsilon_{nmp}\mathcal{A}_p
    \end{equation}
    \begin{eqnarray}
        \omega^{0i}_t=e^{2U}\partial_m U\delta^i_m\nonumber\\
        \omega^{ij}_t=-e^{2U}\delta^i_n\delta^j_m\epsilon_{nmp}\mathcal{A}_p\nonumber\\
        \omega^{0i}_n=-e^{2U}\partial_{[n} U\omega_{m]}\delta^i_m+\delta^i_m\epsilon_{nmp}\mathcal{A}_p\nonumber\\
        \omega^{ij}_n=-\partial_{[m}U\delta^i_{n]}\delta^j_m-e^{2U}\delta^i_p\delta^j_m \epsilon_{pq[m}\omega_{n]}\mathcal{A}_q
    \end{eqnarray}
    \begin{eqnarray}
        \omega^{-0i}_t=\frac{1}{2}(\omega^{0i}_t-\tilde{\omega}^{0i}_t)\nonumber\\
        \tilde{\omega}^{0i}_t=-\frac{i}{2}\epsilon^{0ijk}(\omega_{jk})_t=\frac{i}{2}\epsilon_{ijk}\omega^{jk}_t=\frac{i}{2}e^{2U}\epsilon_{ijk}\delta^j_n\delta^k_m\epsilon_{nmp}\mathcal{A}_p=-ie^{2U}\delta^i_p\mathcal{A}_p\nonumber
    \end{eqnarray}
    \begin{equation}
        \omega^{-0i}_t=\frac{1}{2}e^{2U}\delta^i_m(\partial_m U+i\mathcal{A}_m)
    \end{equation}
    \begin{eqnarray}
        \omega^{-ij}_t=\frac{1}{2}(\omega^{ij}_t-\tilde{\omega}^{ij}_t)\nonumber\\
        \tilde{\omega}^{ij}_t=-i\epsilon_{ijk}\omega^{0k}_t=-ie^{2U}\epsilon_{ijk}\partial_m U\delta^k_m\nonumber\\
        \omega^{-ij}_t=\frac{1}{2}ie^{2U}\epsilon_{ijk}\delta^k_m(\partial_m U+i\mathcal{A}_m)
    \end{eqnarray}
    Solving for the spatial components of the gauge fields one obtains
    \begin{eqnarray}
        F^I_{ij}=e^{2U}\epsilon_{ijk}\delta^k_m\partial_m\tilde{H}^I\nonumber\\
        G_{Iij}=e^{2U}\epsilon_{ijk}\delta^k_m\partial_m H_I
    \end{eqnarray}
    Here, we can use the defining relations for $G_I$ to find the other components, and they are
    \begin{eqnarray}
        F^I_{0i}=e^{2U}\delta^i_m\partial_m(Z^I+\bar{Z}^I)\nonumber\\
        G_{I0i}=e^{2U}\delta^i_m\partial_m(F_I+\bar{F}_I)
    \end{eqnarray}
    \begin{eqnarray}
        T^-_{ij}=2ie^{\mathcal{K}/2}\left(F_IF^{-I}_{ij}-Z^IG^{-}_{Iij}\right)\nonumber\\
        T^-_{ij}=2ie^{3U}\epsilon_{ijk}\delta^k_m\left(F_I\partial_m \tilde{H}^I-Z^I\partial_m H_I\right)
    \end{eqnarray}
    Also,
    \begin{eqnarray}
        \tilde{T}^-_{ab}=-T^-_{ab}\nonumber\\
        T^-_{ab}=\frac{i}{2}\epsilon_{abcd}T^{-cd}\nonumber\\
        T^-_{0i}=\frac{i}{2}\epsilon_{ijk}T^{-jk}\nonumber\\
        T^-_{0i}=-2e^{3U}\delta^i_m\left(F_I\partial_m \tilde{H}^I-Z^I\partial_m H_I\right)
    \end{eqnarray}
    We know that,
    \begin{eqnarray}
        e^{-2U}=e^{-\mathcal{K}}=-i(Z^I\bar{F}_I-F_I\bar{Z}^I)\nonumber\\
        \partial_m U=\frac{1}{2}ie^{2U}(\partial_m Z^I \bar{F}_I+Z^I \partial_m \bar{F}_I-\partial_m F_I\bar{Z}^I-F_I\partial_m \bar{Z}^I)\nonumber\\
        i\mathcal{A}_m=-\frac{1}{2}i e^{2U}\left[(Z^I-\bar{Z}^I)\partial_m(F_I-\bar{F}_I)-(F_I-\bar{F}_I)\partial_m(Z^I-\bar{Z}^I)\right]\nonumber\\
        \partial_m U+i\mathcal{A}_m=e^{2U}(Z^I\partial_m H_I-F_I\partial_m\tilde{H}^I)=\frac{1}{2}e^{-U}\delta^i_m T^{-}_{0i}\nonumber\\
    \end{eqnarray}
    \begin{eqnarray}
        \omega^{-0i}_t=\frac{1}{2}e^{2U}\delta^i_m(\partial_m U+i\mathcal{A}_m)=-\frac{1}{4}e^{U}T^{-0i}\nonumber\\
        \omega^{-ij}_t=\frac{1}{2}ie^{2U}\epsilon_{ijk}\delta^k_m(\partial_m U+i\mathcal{A}_m)=\frac{1}{4}ie^{U}\epsilon_{ijk}T^-_{0k}=-\frac{1}{4}e^UT^{-ij}\nonumber\\
        \omega_m^{-0i}=-\frac{1}{4}e^U\omega_mT^{-0i}+\frac{i}{4}e^{-U}\delta^i_n\epsilon_{nmk}T^{-0k}
    \end{eqnarray}
    finding $G_{ab}^{-\alpha}$,
    \begin{eqnarray}
        \nabla_\alpha X^I G^+_{Iab}-\nabla_\alpha F_I F^{+I}_{ab}=2ig_{\alpha \bar{\beta}}G^{+\bar{\beta}}_{ab}\nonumber\\
        \nabla_\alpha X^I G_{Iab}-\nabla_\alpha F_I F^{I}_{ab}=2ig_{\alpha \bar{\beta}}G^{+\bar{\beta}}_{ab}\nonumber\\
        \epsilon_{ijk}\delta^k_m e^{3U}(\nabla_\alpha Z^I\partial_m H_I-\nabla_\alpha F_I \partial_m \tilde{H}^I)=2ig_{\alpha\bar{\beta}}G^{+\bar{\beta}}_{ij}\nonumber\\
        g_{\alpha \bar{\beta}}G^{+\bar{\beta}}_{ij}=\frac{1}{2i}e^{3U}\epsilon_{ijk}\delta^k_m i(\nabla_\alpha F_I\partial_{\bar{\beta}}\bar{Z}^I-\nabla_\alpha Z^I\partial_{\bar{\beta}}\bar{F}_I)\partial_m \bar{z}^{\bar{\beta}}\nonumber\\
        =-\frac{1}{2}e^{3U}\epsilon_{ijk}\delta^k_m\langle \nabla_\alpha v,\bar{\nabla}_{\bar{\beta}}\bar{v}\rangle\partial_m\bar{z}^{\bar{\beta}}\nonumber\\
        =\frac{i}{2}e^U\epsilon_{ijk}\delta^k_mg_{\alpha\bar{\beta}}\partial_m\bar{z}^{\bar{\beta}}\nonumber\\
        G^{-\beta}_{ij}=-\frac{i}{2}\epsilon_{ijk}\delta^k_m \partial_m z^\beta\nonumber\\
        G^{-\beta}_{0i}=\frac{i}{2}\epsilon_{ijk}G^{-\beta jk}\nonumber\\
        G^{-\beta}_{0i}=\frac{1}{2}e^U\delta^i_m\partial_m z^\beta
    \end{eqnarray}
    \section{Equations of motion for the fermions}
    \label{app:eomf}
     expanding \ref{eq:fullAction} we get
    \begin{eqnarray}
        -\bar{\psi}_{i a}\gamma^{a b c}D_b\psi_c^i+\Big(F^{+I}_{a b}Im{\mathcal{N}}_{IJ}{\nabla}_{{\alpha}}{X}^J\Big(\frac{1}{8}g^{{\alpha}\bar{\beta}}C_{\bar{\beta}\bar{\gamma}\bar{\delta}}\bar{\chi}^{i\bar{\gamma}}\gamma_{ab}\chi^{j\bar{\delta}}\varepsilon_{ij}+\bar{\chi}^{{\alpha}}_i\gamma^a\psi^{b}_j\varepsilon^{ij}\Big)\nonumber\\
        +F^{+I}_{a b}Im{\mathcal{N}}_{IJ}\bar{X}^J\bar{\psi}^{ai}\psi^{bj}\varepsilon_{ij}-\frac{1}{4}g_{\alpha\bar{\beta}}\bar{\chi}_i^\alpha \slashed{D}\chi^{i\bar{\beta}}+\frac{1}{2}g_{\alpha\bar{\beta}}\bar{\psi}_{ia}\slashed{\partial}z^\alpha\gamma^a\chi^{i\bar{\beta}}\nonumber\\
        +F^{-}_{ab}Im\mathcal{N}_{IJ} \bar{\nabla}_{\bar{\alpha}}\bar{X}^J\Big(\frac{1}{8}g^{\beta\bar{\alpha}}C_{\beta\gamma\delta}\bar{\chi}_i^\gamma\gamma_{ab}\chi_j^\delta\varepsilon^{ij}+\bar{\chi}^{\bar{\alpha}i}\gamma^a\psi^{bj}\varepsilon_{ij}\Big)\nonumber\\
        +F^-_{ab}Im\mathcal{N}_{IJ}X^J\bar{\psi}^a_i\psi^b_j\varepsilon^{ij}-\frac{1}{4}g_{\bar{\alpha}{\beta}}\bar{\chi}^{i\bar{\alpha}} \slashed{D}\chi^{{\beta}}_i+\frac{1}{2}g_{\bar{\alpha}{\beta}}\bar{\psi}^i_{a}\slashed{\partial}\bar{z}^{\bar{\alpha}}\gamma^a\chi^{{\beta}}_i\Big)
    \end{eqnarray}
    Now varying the action w.r.t. fermions to find the equations of motion of the fermions we get
    \begin{eqnarray}
        e^{-1}\delta\mathcal{L}_{fermi}=-\delta\bar{\psi}_{i a}\gamma^{a b c}D_b\psi_c^i-\bar{\psi}_{i a}\gamma^{a b c}D_b\delta\psi_c^i\nonumber\\
        +F^{+I a b}Im{\mathcal{N}}_{IJ}{\nabla}_{{\alpha}}{X}^J\Big(\frac{1}{8}g^{{\alpha}\bar{\beta}}C_{\bar{\beta}\bar{\gamma}\bar{\delta}}\delta\bar{\chi}^{i\bar{\gamma}}\gamma_{ab}\chi^{j\bar{\delta}}\varepsilon_{ij}+\frac{1}{8}g^{{\alpha}\bar{\beta}}C_{\bar{\beta}\bar{\gamma}\bar{\delta}}\bar{\chi}^{i\bar{\gamma}}\gamma_{ab}\delta\chi^{j\bar{\delta}}\varepsilon_{ij}\nonumber\\
        +\delta\bar{\chi}^{{\alpha}}_i\gamma_a\psi_{bj}\varepsilon^{ij}
        +\bar{\chi}^{{\alpha}}_i\gamma_a\delta\psi_{bj}\varepsilon^{ij}\Big)+F^{+I}_{a b}Im{\mathcal{N}}_{IJ}\bar{X}^J\delta\bar{\psi}^{ai}\psi^{bj}\varepsilon_{ij}+F^{+I}_{a b}Im{\mathcal{N}}_{IJ}\bar{X}^J\bar{\psi}^{ai}\delta\psi^{bj}\varepsilon_{ij}\nonumber\\
        -\frac{1}{4}g_{\alpha\bar{\beta}}\delta\bar{\chi}_i^\alpha \slashed{D}\chi^{i\bar{\beta}}-\frac{1}{4}g_{\alpha\bar{\beta}}\bar{\chi}_i^\alpha \slashed{D}\delta\chi^{i\bar{\beta}}+\frac{1}{2}g_{\alpha\bar{\beta}}\delta\bar{\psi}_{ia}\slashed{\partial}z^\alpha\gamma^a\chi^{i\bar{\beta}}+\frac{1}{2}g_{\alpha\bar{\beta}}\bar{\psi}_{ia}\slashed{\partial}z^\alpha\gamma^a\delta\chi^{i\bar{\beta}}\nonumber\\
        +F^{-ab}Im\mathcal{N}_{IJ} \bar{\nabla}_{\bar{\alpha}}\bar{X}^J\Big(\frac{1}{8}g^{\beta\bar{\alpha}}C_{\beta\gamma\delta}\delta\bar{\chi}_i^\gamma\gamma_{ab}\chi_j^\delta\varepsilon^{ij}+\frac{1}{8}g^{\beta\bar{\alpha}}C_{\beta\gamma\delta}\bar{\chi}_i^\gamma\gamma_{ab}\delta\chi_j^\delta\varepsilon^{ij}\nonumber\\
        +\delta\bar{\chi}^{\bar{\alpha}i}\gamma_a\psi^{j}_b\varepsilon_{ij}+\bar{\chi}^{\bar{\alpha}i}\gamma_a\delta\psi^{j}_b\varepsilon_{ij}\Big)+F^{-I}_{ab}Im\mathcal{N}_{IJ}X^J\delta\bar{\psi}^a_i\psi^b_j\varepsilon^{ij}+F^{-I}_{ab}Im\mathcal{N}_{IJ}X^J\bar{\psi}^a_i\delta\psi^b_j\varepsilon^{ij}\nonumber\\
        -\frac{1}{4}g_{\bar{\alpha}{\beta}}\delta\bar{\chi}^{i\bar{\alpha}} \slashed{D}\chi^{{\beta}}_i-\frac{1}{4}g_{\bar{\alpha}{\beta}}\bar{\chi}^{i\bar{\alpha}} \slashed{D}\delta\chi^{{\beta}}_i+\frac{1}{2}g_{\bar{\alpha}{\beta}}\delta\bar{\psi}^i_{a}\slashed{\partial}\bar{z}^{\bar{\alpha}}\gamma^a\chi^{{\beta}}_i+\frac{1}{2}g_{\bar{\alpha}{\beta}}\bar{\psi}^i_{a}\slashed{\partial}\bar{z}^{\bar{\alpha}}\gamma^a\delta\chi^{{\beta}}_i
    \end{eqnarray}
     collecting terms with $\delta\bar{\psi}_{ia}$
    \begin{eqnarray}
        -\delta{\bar{\psi}}_{ia}\gamma^{abc}D_b\psi_c^i+F^{+Iab}Im\mathcal{N}_{IJ}\nabla_\alpha X^J\bar{\chi}_i^\alpha\gamma_a\delta\psi_{jb}\varepsilon^{ij}+\frac{1}{2}g_{\alpha\bar{\beta}}\delta\bar{\psi}_{ia}\slashed{\partial}z^\alpha\gamma^a\chi^{i\bar{\beta}}\nonumber\\
        +F^{-Iab}Im\mathcal{N}_{IJ}X^J\delta\bar{\psi}_{ia}\psi_{jb}\varepsilon^{ij}+F^{-Iab}Im\mathcal{N}_{IJ}X^J\bar{\psi}_{ia}\delta\psi_{jb}\varepsilon^{ij}
    \end{eqnarray}
    using spinor flip properties $\bar{\lambda}\gamma_{\mu_1\cdots\mu_r}\chi=t_r\bar{\chi}\gamma_{\mu_1\cdots\mu_r}\lambda$
    \begin{eqnarray}
        -\delta{\bar{\psi}}_{ia}\gamma^{abc}D_b\psi_c^i-F^{+Iab}Im\mathcal{N}_{IJ}\nabla_\alpha X^J\delta\bar{\psi}_{ia}\gamma_b\chi_{j}^\alpha\varepsilon^{ij}+\frac{1}{2}g_{\alpha\bar{\beta}}\delta\bar{\psi}_{ia}\slashed{\partial}z^\alpha\gamma^a\chi^{i\bar{\beta}}\nonumber\\
        +2F^{-Iab}Im\mathcal{N}_{IJ}X^J\delta\bar{\psi}_{ia}\psi_{jb}\varepsilon^{ij}
    \end{eqnarray}
    Hence one of the equations of motion is
    \begin{eqnarray}
        -\gamma^{abc}D_b\psi_c^i-F^{+Iab}Im\mathcal{N}_{IJ}\nabla_\alpha X^J\gamma_b\chi_j^\alpha\varepsilon^{ij}+\frac{1}{2}g_{\alpha\bar{\beta}}\slashed{\partial}z^\alpha\gamma^a\chi^{i\bar{\beta}}+2F^{-Iab}Im\mathcal{N}_{IJ}X^J\psi_{jb}\varepsilon^{ij}=0\nonumber\\
    \end{eqnarray}
    from the definitions
    \begin{eqnarray}
        G^{-\alpha ab}=g^{\alpha\bar{\beta}} F^{-I ab}Im\mathcal{N}_{IJ}\bar{\nabla}_{\bar{\beta}}\bar{X}^J\nonumber\\
        T^+_{ab}=-4F^{+I}_{ab}Im\mathcal{N}_{IJ}\bar{X}^J
    \end{eqnarray}
    The equations of motion for the gravitini can be written as
    \begin{eqnarray}
        \frac{\delta{\mathcal{L}}}{\delta\bar{\psi}_{ia}}=-\gamma^{abc}D_b\psi_c^i-F^{+Iab}Im\;\mathcal{N}_{IJ}\nabla_\alpha X^J\gamma_b\chi_j^\alpha\varepsilon^{ij}+2F^{-Iab}Im\;\mathcal{N}_{IJ}X^J\psi_{jb}\varepsilon^{ij}\nonumber\\
        +\frac{1}{2}g_{\alpha\bar{\beta}}\slashed{\partial}z^\alpha\gamma^a\chi^{i\bar{\beta}}=0\nonumber\\
        -\gamma^{abc}D_b\psi_c^i-g_{\alpha\bar{\beta}}G^{+\bar{\beta}ab}\gamma_b\chi^\alpha_j\varepsilon^{ij}-\frac{1}{2}T^{-ab}(\psi_{j})_b\varepsilon^{ij}+\frac{1}{2}g_{\alpha\bar{\beta}}\slashed{\partial}z^\alpha\gamma^a\chi^{i\bar{\beta}}=0\nonumber\\
    \end{eqnarray}
    now we collect the terms containing $\delta \chi^{i\bar{\alpha}}$
    \begin{eqnarray}
        F^{+I a b}Im{\mathcal{N}}_{IJ}{\nabla}_{{\alpha}}{X}^J\Big(\frac{1}{8}g^{{\alpha}\bar{\beta}}C_{\bar{\beta}\bar{\gamma}\bar{\delta}}\delta\bar{\chi}^{i\bar{\gamma}}\gamma_{ab}\chi^{j\bar{\delta}}\varepsilon_{ij}+\frac{1}{8}g^{{\alpha}\bar{\beta}}C_{\bar{\beta}\bar{\gamma}\bar{\delta}}\bar{\chi}^{i\bar{\gamma}}\gamma_{ab}\delta\chi^{j\bar{\delta}}\varepsilon_{ij}\Big)\nonumber\\
        -\frac{1}{4}g_{\alpha\bar{\beta}}\bar{\chi}_i^\alpha \slashed{D}\delta\chi^{i\bar{\beta}}+\frac{1}{2}g_{\alpha\bar{\beta}}\bar{\psi}_{ia}\slashed{\partial}z^\alpha\gamma^a\delta\chi^{i\bar{\beta}}+F^{-ab}Im\mathcal{N}_{IJ} \bar{\nabla}_{\bar{\alpha}}\bar{X}^J\delta\bar{\chi}^{\bar{\alpha}i}\gamma_a\psi^{j}_b\varepsilon_{ij}\nonumber\\
        -\frac{1}{4}g_{\bar{\alpha}{\beta}}\delta\bar{\chi}^{i\bar{\alpha}} \slashed{D}\chi^{{\beta}}_i\nonumber
    \end{eqnarray}
    \begin{eqnarray}
        -\frac{1}{4}g_{\alpha\bar{\beta}}\bar{\chi}_i^\alpha \slashed{D}\delta\chi^{i\bar{\beta}}=-\frac{1}{4}g_{\alpha\bar{\beta}}\bar{\chi}_i^\alpha\left(\gamma^\mu\partial_\mu+\frac{1}{4}\gamma^\mu\omega_\mu^{ab}\gamma_{ab}-\frac{i}{2}\gamma^\mu\mathcal{A}_\mu\right)\delta\chi^{i\bar{\beta}}-\frac{1}{4}g_{\alpha\bar{\beta}}\bar{\chi}_i^\alpha\gamma^\mu\Gamma^{\bar{\beta}}_{\bar{\delta}\bar{\gamma}}\partial_\mu\bar{z}^{\bar{\delta}}\delta\chi^{i\bar{\gamma}}\nonumber\\
        =-\frac{1}{4}g_{\alpha\bar{\beta}}\left(-\partial_\mu\bar{\delta{\chi}}^{i\bar{\beta}}\gamma^\mu\chi_i^\alpha+\frac{1}{4}\omega_\mu^{ab}\delta{\chi}^{i\bar{\beta}}\gamma_{ab}\gamma^\mu\chi_i^\alpha+\frac{i}{2}\delta\chi^{i\bar{\beta}}\gamma^\mu\mathcal{A}_\mu\chi_i^\alpha\right)\delta\chi^{i\bar{\beta}}+\frac{1}{4}g_{\alpha\bar{\beta}}\delta\bar{\chi}^{i\bar{\gamma}}\gamma^\mu\Gamma^{\bar{\beta}}_{\bar{\delta}\bar{\gamma}}\partial_\mu\bar{z}^{\bar{\delta}}\chi_i^\alpha\nonumber\\
        =-\frac{1}{4}g_{\alpha\bar{\beta}}\left({\delta\bar{\chi}}^{i\bar{\beta}}\gamma^\mu\partial_\mu\chi_i^\alpha+\frac{1}{4}\omega_\mu^{ab}\delta\bar{\chi}^{i\bar{\beta}}\gamma^\mu\gamma_{ab}\chi_i^\alpha+\frac{i}{2}\delta\bar{\chi}^{i\bar{\beta}}\gamma^\mu\mathcal{A}_\mu\chi_i^\alpha\right)+\frac{1}{4}g_{\alpha\bar{\beta}}\delta\bar{\chi}^{i\bar{\gamma}}\gamma^\mu\Gamma^{\bar{\beta}}_{\bar{\delta}\bar{\gamma}}\partial_\mu\bar{z}^{\bar{\delta}}\chi_i^\alpha\nonumber\\
    \end{eqnarray}
    \begin{equation}
        \left(\partial_\mu e^a_\nu+\left(\omega^{a}_{b}\right)_\mu e^b_\nu\right) dx^\mu\wedge dx^\nu=0
    \end{equation}
    \begin{eqnarray}
        =-\frac{1}{4}g_{\alpha\bar{\beta}}\left({\delta\bar{\chi}}^{i\bar{\beta}}\gamma^\mu\partial_\mu\chi_i^\alpha+\frac{1}{4}\omega_\mu^{ab}\delta\bar{\chi}^{i\bar{\beta}}\gamma^\mu\gamma_{ab}\chi_i^\alpha+\frac{i}{2}\delta\bar{\chi}^{i\bar{\beta}}\gamma^\mu\mathcal{A}_\mu\chi_i^\alpha\right)\nonumber\\
        -\frac{1}{4}g_{\alpha\bar{\beta}}\delta\bar{\chi}^{i\bar{\beta}}\Gamma^\alpha_{\delta\gamma}\gamma^\mu\chi^\gamma_i\partial_\mu z^\delta+\frac{1}{4}g_{\alpha\bar{\beta}}\delta\bar{\chi}^{i\bar{\gamma}}\gamma^\mu\Gamma^{\bar{\beta}}_{\bar{\delta}\bar{\gamma}}\partial_\mu\bar{z}^{\bar{\delta}}\chi_i^\alpha\nonumber\\
        +\frac{1}{4}g_{\alpha\bar{\beta}}\delta\bar{\chi}^{i\bar{\beta}}\Gamma^\alpha_{\delta\gamma}\gamma^\mu\chi^\gamma_i\partial_\mu z^\delta\nonumber\\
        =-\frac{1}{4}g_{\alpha\bar{\beta}}\delta\bar{\chi}^{i\bar{\beta}}\slashed{D}\chi_i^\alpha+\frac{1}{4}g_{\alpha\bar{\beta}}\delta\bar{\chi}^{i\bar{\gamma}}\gamma^\mu\Gamma^{\bar{\beta}}_{\bar{\delta}\bar{\gamma}}\partial_\mu\bar{z}^{\bar{\delta}}\chi_i^\alpha+\frac{1}{4}g_{\alpha\bar{\beta}}\delta\bar{\chi}^{i\bar{\beta}}\Gamma^\alpha_{\delta\gamma}\gamma^\mu\chi^\gamma_i\partial_\mu z^\delta
    \end{eqnarray}
    \begin{eqnarray}
        \frac{1}{4}g_{\alpha\bar{\beta}}\delta\bar{\chi}^{i\bar{\gamma}}\gamma^\mu\Gamma^{\bar{\beta}}_{\bar{\delta}\bar{\gamma}}\partial_\mu\bar{z}^{\bar{\delta}}\chi_i^\alpha+\frac{1}{4}g_{\alpha\bar{\beta}}\delta\bar{\chi}^{i\bar{\beta}}\Gamma^\alpha_{\delta\gamma}\gamma^\mu\chi^\gamma_i\partial_\mu z^\delta\nonumber\\=\frac{1}{4}\delta\chi^{i\bar{\beta}}\gamma^\mu\chi_i^\alpha(\partial_\delta g_{\alpha\bar{\beta}}\partial_\mu z^\delta+\partial_{\bar{\delta}}g_{\alpha\bar{\beta}}\partial_\mu\bar{z}^{\bar{\delta}})=\frac{1}{4}\delta\chi^{i\bar{\beta}}\gamma^\mu\chi_i^\alpha\partial_\mu g_{\alpha\bar{\beta}}
    \end{eqnarray}
    \begin{eqnarray}
        -\frac{1}{4}g_{\alpha\bar{\beta}}\bar{\chi}_i^\alpha \slashed{D}\delta\chi^{i\bar{\beta}}=-\frac{1}{4}g_{\alpha\bar{\beta}}\delta\bar{\chi}^{i\bar{\beta}}\slashed{D}\chi_i^\alpha+\frac{1}{4}\delta\chi^{i\bar{\beta}}\gamma^\mu\chi_i^\alpha\partial_\mu g_{\alpha\bar{\beta}}
    \end{eqnarray}
    Now, the terms of the action containing $\delta\chi^{i\bar{\alpha}}$ can be written as
    \begin{eqnarray}
        F^{+I a b}Im{\mathcal{N}}_{IJ}{\nabla}_{{\alpha}}{X}^J\Big(\frac{1}{8}g^{{\alpha}\bar{\beta}}C_{\bar{\beta}\bar{\gamma}\bar{\delta}}\delta\bar{\chi}^{i\bar{\gamma}}\gamma_{ab}\chi^{j\bar{\delta}}\varepsilon_{ij}+\frac{1}{8}g^{{\alpha}\bar{\beta}}C_{\bar{\beta}\bar{\gamma}\bar{\delta}}\delta\bar{\chi}^{i\bar{\delta}}\gamma_{ab}\chi^{j\bar{\gamma}}\varepsilon_{ij}\Big)\nonumber\\
        +\frac{1}{4}\delta\chi^{i\bar{\beta}}\gamma^\mu \chi_i^\alpha\partial_\mu g_{\alpha\bar{\beta}}+\frac{1}{2}g_{\alpha\bar{\beta}}\bar{\psi}_{ia}\slashed{\partial}z^\alpha\gamma^a\delta\chi^{i\bar{\beta}}+F^{-ab}Im\mathcal{N}_{IJ} \bar{\nabla}_{\bar{\alpha}}\bar{X}^J\delta\bar{\chi}^{\bar{\alpha}i}\gamma_a\psi^{j}_b\varepsilon_{ij}\nonumber\\
        -\frac{1}{2}g_{\bar{\alpha}{\beta}}\delta\bar{\chi}^{i\bar{\alpha}} \slashed{D}\chi^{{\beta}}_i\nonumber
    \end{eqnarray}
    The equation of motion for the gaugino then is
    \begin{eqnarray}
        \frac{1}{4}F^{+I a b}Im{\mathcal{N}}_{IJ}{\nabla}_{{\alpha}}{X}^Jg^{{\alpha}\bar{\gamma}}C_{\bar{\gamma}\bar{\beta}\bar{\delta}}\gamma_{ab}\chi^{j\bar{\delta}}\varepsilon_{ij}+\frac{1}{4}\gamma^\mu \chi_i^\alpha\partial_\mu g_{\alpha\bar{\beta}}+\frac{1}{2}g_{\alpha\bar{\beta}}\gamma^a\slashed{\partial}z^\alpha{\psi}_{ia}\nonumber\\
        +F^{-Iab}Im\mathcal{N}_{IJ} \bar{\nabla}_{\bar{\beta}}\bar{X}^J\gamma_a\psi^j_b\varepsilon_{ij}-\frac{1}{2}g_{\alpha\bar{\beta}}\slashed{D}\chi_i^{\alpha}=0.
    \end{eqnarray}
    \bibliographystyle{apsrev4-1}
    \bibliography{ref}
\end{document}